\documentclass[%
 reprint,
 amsmath,amssymb,
pra,
]{revtex4-2}
\usepackage{booktabs}
\usepackage{multirow}
\usepackage{graphicx}% Include figure files
\usepackage{dcolumn}% Align table columns on decimal point
\usepackage{bm}% bold math
\begin{document}

\preprint{APS/123-QED}

\title{ Fully passive quantum key distribution with parametric down-conversion source }% Force line breaks with \\
\author{Jia-Wei Ying,$^{1}$ Qi Zhang,$^{2}$  Shi-Pu Gu,$^{1}$ Xing-Fu Wang,$^{2}$ Lan Zhou,$^{2}$}\email{zhoul@njupt.edu.cn}
\author{Yu-Bo Sheng$^{1}$}
 \email{shengyb@njupt.edu.cn}
\affiliation{%
 $^1$College of Electronic and Optical Engineering and College of Flexible Electronics (Future Technology), Nanjing
 University of Posts and Telecommunications, Nanjing, 210023, China\\
 $^2$College of Science, Nanjing University of Posts and Telecommunications, Nanjing, 210023, China\\
}%

\date{\today}% It is always \today, today,

\begin{abstract}
The fully passive source is capable of passively generating decoy states and performing passive encoding simultaneously, avoiding the side-channel risks caused by active modulation operations at the source end, thus effectively enhance the security in quantum key distribution (QKD). Existing fully passive QKD protocol and experiments exploit phase-randomized coherent pulses. In this paper, we propose a fully passive QKD protocol using  parametric down-conversion source. The decoy state generation and encoding operation can be carried out passively by parameter down-conversion progress. This protocol has several advantages. First, it can also eliminate all side channels in
active modulators. Second, compared with fully passive QKD protocol with phase-randomized coherent pulses, our protocol can significantly increase the key rate and extend the communication distance. Meanwhile, in terms of the transmission rate, our protocol is also closer to that of actively modulated QKD and can achieve fully passive modulation with fewer resources. Moreover, combined with measurement-device-independent (MDI) QKD,  this protocol can even potentially achieve robustness against side channels in both detectors and modulators.

\end{abstract}

\maketitle

%\tableofcontents

\section{Introduction}
Quantum key distribution (QKD) enables two participants to share a secure key  \cite{bb84}. Theoretically, QKD offers  unconditional security. However, existing technologies fall short of fulfilling the stringent security requirements of QKD in practical scenarios, leaving multiple side channels exposed to potential eavesdroppers \cite{QHACK1,QHACK2,QHACK3,QHACK4,QHACK5}.  For instance, at the source end, active modulation operations are typically required, including active encoding and the active preparation of  decoy states. These active modulation processes may expose certain side channels, providing opportunities for eavesdroppers to execute Trojan horse attacks \cite{trojan0,trojan1,trojan2,trojan3,trojan4,trojan5,trojan6}.

Passive sources is an effective means to mitigate side channels at the  source endpoint.
Over the past few decades, passive encoding \cite{passivestate1}  and passive decoy state \cite{pdecoy}   protocols based on coherent pulse have been proposed and implemented in various protocols \cite{passivestatecv1,passivestatecv2,passivestatecv3,pdecoy1,pdecoy2,pdecoy3,pdecoy4,pdecoy5,pdecoy6,pdecoy7}. In 2022, Wang et. al. introduced a fully passive protocol that combines passive encoding and passive decoy states \cite{FP1}, utilizing phase-randomized coherent pulses to eliminate all side channels at the light source end. The protocol have been incorporated into many quantum communication protocols\cite{FP3,FP4,FPMDI1,FPMDI2,FP5,FPTW,FPCKA} and have been experimentally validated\cite{FPexp1,FPexp2}.

In addition to schemes based on coherent pulses, there is another approach for achieving passive decoy states in passive source configurations, which is based on parametric down-conversion (PDC) source \cite{pdecoypdc1}. These protocols leverage heralding operations to produce sources with excellent pulse performance. Over the past decade, PDC-based passive decoy state protocols have been utilized in various protocols \cite{pdecoypdc2,pdecoypdc3,pdecoypdc4,pdecoypdc5,pdecoypdc6,pdecoypdc7}. However, to date, no passive encoding protocol based on PDC has been proposed.

Inspired by the previous work \cite{pdecoypdc1,pdecoypdc2,pdecoypdc3,pdecoypdc4,pdecoypdc5,pdecoypdc6,pdecoypdc7}, we design a herald fully passive light source based on PDC and apply it to QKD protocols.  This method mainly has the following advantages.
Firstly, through the heralding operation, the passive generation of decoy states and passive encoding can be achieved synchronously.
This eliminates the need for active modulation and consequently wipes out  all side channels present at the source end. Secondly, the heralding operation can effectively reduce the vacuum state component in the source, thereby increasing the proportion of single-photon component and significantly improving the communication quality.
Compared to actively modulated QKD, our protocol demonstrates significant advantages in key rate and transmission distance.
Thirdly, we have optimized the key transmission rate, achieving higher performance than fully passive protocols based on coherent light.
Furthermore, our herald fully passive  source can be integrated with measurement-device-independent (MDI) protocols, affording the potential to further nullify side channels at the detector end and enhance the security of QKD.

The structure of the paper is as follows. In Sec. II, we introduce  system model, including the herald fully passive source, channel model, key generation rate, and decoy state analysis. In Sec. III, we conduct numerical simulation and parameter optimization. In Sec. IV, we give a  conclusion.

\section{System model }
\subsection{The herald fully passive source} \label{sa}
\begin{figure}[!htbp]%[tpb]
	\begin{center}
		
		\includegraphics[width=8.5cm,angle=0]{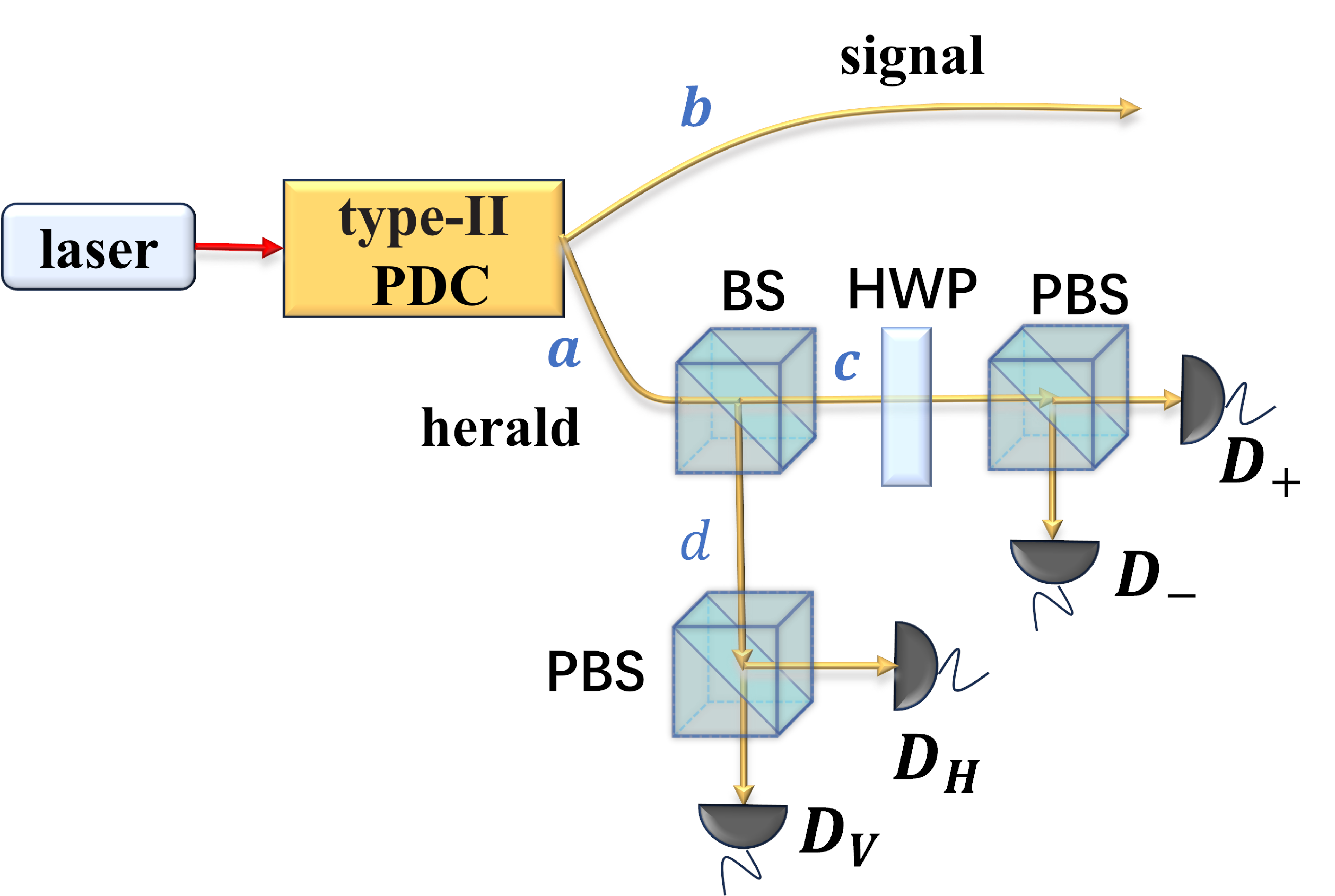}
		%\captionsetup{justification=raggedright}
		\caption{The structure of the herald fully passive source. BS: beam splitter; PBS: polarization beam splitter; HWP: half wave plate; $D_{+},D_{-},D_{H},D_{V}$: detector.  }\label{pro}
	\end{center}
\end{figure}

In this section, we will introduce the herald fully passive source, whose structure is illustrated in Fig. \ref{pro}. Through post-selection, it can perform passive encoding operations and passively generate decoy states, simultaneously.

The source is composed of a PDC segment and a herald measurement segment. Initially, a laser beam is made incident upon a type-II PDC crystal, thereby generating  entangled-photon pairs.
The quantum state engendered thereby can be formulated as \cite{source1,source2}:
\begin{eqnarray}
	| \psi   \rangle =\frac{1}{cosh^{2}\chi } \sum_{n=0}^{\infty } \sqrt{n+1}  tanh^{n}\chi| \Phi _{n}   \rangle, \label{sour}
\end{eqnarray}
where $\chi$ is the squeezing coefficient and  $| \Phi _{n}   \rangle $ is the state of an $n$-photon pair, given by:
\begin{eqnarray}
	|   \Phi _{n} \rangle  =\frac{1}{\sqrt{n+1} } \sum_{m=0}^{n} |   H \rangle _{a}^{n-m}
	|   V \rangle _{a}^{m}|   H \rangle _{b}^{m}|   V \rangle _{b}^{n-m},
\end{eqnarray}
where the subscript $a$ $(b)$ represent the heralding (signal) path, and $|   H \rangle$ $(|   V \rangle)$ represents the horizontal (vertical) polarization of photons.
Note that when n is equal to 1, the single-photon pair will be a Bell state:
\begin{eqnarray}\label{bell}
	|   \Phi _{1} \rangle  &&=\frac{1}{\sqrt{2} }
	(|   H \rangle _{a}|   V \rangle _{b}+|   V \rangle _{a}|   H \rangle _{b}) \\ \nonumber
	&&=\frac{1}{\sqrt{2} }
	(|   + \rangle _{a}|   + \rangle _{b}-|   - \rangle _{a}|   - \rangle _{b}).
\end{eqnarray}

The underlying principle of our protocol is fairly simple. In the context of an ideal pair of entangled photons, we detect the photon in the heralding path. When the detector $D_{H}$ gives a response, it indicates that the photon in the signal path is $|   V \rangle$; likewise, if the detector $D_{+}$ responds, it means that the photon in the signal path is $|   + \rangle$. By employing this approach, passive encoding can be accomplished. By means of different response combinations of the detectors, we can generate decoy states with diverse photon distributions.  Next, let's take into account the light source that would actually generate multiple photon pairs.

In Eq. (\ref{sour}), there exists a variable $\chi $, which represents the squeezing coefficient. To begin with, we'll transform it into the average photon pair number that is more commonly recognized.
Here, we define $\lambda =sinh^{2}\chi$ and the average photon pair number of the source can be calculated as:
\begin{eqnarray}
	\bar{n} &&= \sum_{n=0}^{\infty }  (\frac{1}{cosh^{2}\chi } \sqrt{n+1}  tanh^{n}\chi  ) ^{2}*n \\ \nonumber
	&&=\sum_{n=0}^{\infty } n(n+1)(\frac{\lambda }{\lambda+1} )^{n}\frac{1}{\lambda+1} \\ \nonumber
	&&\overset{\frac{\lambda }{\lambda+1}=x}{\longrightarrow } \frac{1}{(\lambda+1)^{2}}
	\sum_{n=0}^{\infty }n(n+1)x^{n}\\ \nonumber
	&&=\frac{1}{(\lambda+1)^{2}}\sum_{n=0}^{\infty }x\frac{d^{2}}{dx^{2}} x^{n+1}\\ \nonumber
	&&=\frac{1}{(\lambda+1)^{2}}x\frac{d^{2}}{dx^{2}}(\sum_{n=0}^{\infty }x^{n+1})\\ \nonumber
	&&=\frac{1}{(\lambda+1)^{2}}x\frac{d^{2}}{dx^{2}}\frac{x}{1-x} \\ \nonumber
	&&=\frac{1}{(\lambda+1)^{2}}2x(1-x)^{-3}\\ \nonumber
	&&=2\lambda,
\end{eqnarray}
and the probability to generate n-photon pair state $|   \Phi _{n} \rangle$ will be
\begin{eqnarray}
	P(n) & = & \frac{(n+1)\lambda ^{n}}{(1+\lambda )^{n+2}} .
\end{eqnarray}

Subsequently, our attention is directed towards the photons within the heralding path, which are randomly chosen to undergo measurement in either the X basis or the Z basis. The photon in the heralding path initially traverses a beam splitter (BS). Of the two outputs of the BS, one passes through a half-wave plate (HWP) and is subsequently fed into a polarizing beam splitter (PBS) for $X$-basis measurement, whereas the other is directly input into the PBS for $Z$-basis measurement.
We are capable of deducing the photon state within the signal path by relying on the response of the heralding path. For instance, in the case where the detection outcome of the heralding path is $ |   H \rangle$, it is deemed that the state emitted from the signal path is $ |   V \rangle$. Likewise, when the detection result of the heralding path is $ |   + \rangle$, it is regarded that the state emitted by the signal path is $ |   + \rangle$.
The derivation of this process is as follows:
\begin{widetext}
\begin{eqnarray}
	|   \Phi _{n} \rangle && =\frac{1}{\sqrt{n+1} } \sum_{m=0}^{n} |   H \rangle _{a}^{n-m}
	|   V \rangle _{a}^{m}|   H \rangle _{b}^{m}|   V \rangle _{b}^{n-m} \nonumber\\\nonumber
	&&=\frac{1}{\sqrt{n+1} } \sum_{m=0}^{n} \frac{1}{(n-m)!m!}
	(a_{H}^{\dagger })^{n-m} (a_{V}^{\dagger })^{m}
	(b_{H}^{\dagger })^{m} (b_{V}^{\dagger })^{n-m} |   vac \rangle \\\nonumber
	&&\overset{BS}{\longrightarrow }\frac{1}{\sqrt{n+1} } \sum_{m=0}^{n} \frac{1}{(n-m)!m!}
	(b_{H}^{\dagger })^{m} (b_{V}^{\dagger })^{n-m}
	  (\frac{c_{H}^{\dagger }+d_{H}^{\dagger }}{\sqrt{2} } )^{n-m}
	(\frac{c_{V}^{\dagger }+d_{V}^{\dagger }}{\sqrt{2} } )^{m} |   vac \rangle \\\nonumber
	&&\overset{HWP}{\longrightarrow }\frac{1}{\sqrt{n+1} } \sum_{m=0}^{n} \frac{1}{(n-m)!m!}
	(b_{H}^{\dagger })^{m} (b_{V}^{\dagger })^{n-m}
	(\frac{1}{2}c_{H}^{\dagger }+\frac{1}{2}c_{V}^{\dagger } +\frac{1}{\sqrt{2} }d_{H}^{\dagger } )^{n-m}
	(\frac{1}{2}c_{H}^{\dagger }-\frac{1}{2}c_{V}^{\dagger }
	+\frac{1}{\sqrt{2} }d_{V}^{\dagger } )^{m} |   vac \rangle \\\nonumber
	&&=\frac{1}{\sqrt{n+1} } \sum_{m=0}^{n}
\sum_{i_{1}+i_{2}+i_{3}=n-m} \sum_{j_{1}+j_{2}+j_{3}=m}   \frac{1}{i_{1}!i_{2}!i_{3}!}
	\frac{1}{j_{1}!j_{2}!j_{3}!} (\frac{1}{\sqrt{2} }) ^{2n-i_{3}-j_{3}}(-1)^{j_{2}}
	\sqrt{(i_{1}+j_{1})!} \sqrt{(i_{2}+j_{2})!} \sqrt{i_{3}!} \sqrt{j_{3}!} \\
	&& \quad\quad \quad\quad| i_{1}+j_{1}  \rangle_{D_{+}} | i_{2}+j_{2}  \rangle_{D_{-}}
	| i_{3}  \rangle_{D_{H}} | j_{3}  \rangle_{D_{V}} \sqrt{m!(n-m)!} | m  \rangle_{b_{H}}| n-m  \rangle_{b_{V}} \\\nonumber
\end{eqnarray}
\end{widetext}

Then we set
\begin{eqnarray}
	i_{1}+j_{1}=n_{+},i_{3}=n_{H}, \\ \nonumber
	i_{2}+j_{2}=n_{-},j_{3}=n_{V},
\end{eqnarray}
and the probability to generate the state $|  n_{+} \rangle_{D_{+}} $ $| n_{-}  \rangle_{D_{-}} $ $| n_{H}  \rangle_{D_{H}}$ $ | n_{V} \rangle_{D_{V}}$ $| m  \rangle_{b_{H}}$ $|n-m\rangle_{b_{V}} $ can be written as:
\begin{eqnarray}
	&&P_{s}(n_{+},n_{-},n_{H},n_{V},m,n-m)=\\\nonumber
	&&\quad\quad P(n) |  \frac{1}{\sqrt{n+1} }
	\sqrt{m!(n-m)!}
	\sum_{i_{1}+j_{1}=n_{+}}\sum_{i_{2}+j_{2}=n_{-}} \\\nonumber
	&&\quad\quad \frac{1}{i_{1}!i_{2}!i_{3}!}
	\frac{1}{j_{1}!j_{2}!j_{3}!}
	 (\frac{1}{\sqrt{2} }) ^{2n-i_{3}-j_{3}}(-1)^{j_{2}}
	\sqrt{(i_{1}+j_{1})!} \\\nonumber
	&&\quad\quad \sqrt{(i_{2}+j_{2})!} \sqrt{i_{3}!} \sqrt{j_{3}!}| ^{2}.
\end{eqnarray}

The state before measurement can be written as:
\begin{eqnarray}
	| \psi_{1}   \rangle=&&\sum_{n=0}^{\infty } \sum_{m=0}^{n}
	\sqrt{P_{s}(n_{+},n_{-},n_{H},n_{V},m,n-m)} \\\nonumber
	&&|  n_{+} \rangle_{D_{+}} | n_{-}  \rangle_{D_{-}} | n_{H}  \rangle_{D_{H}} | n_{V} \rangle_{D_{V}}| m  \rangle_{b_{H}}|n-m\rangle_{b_{V}} .
\end{eqnarray}

Subsequently, the photons within the heralding path are subjected to measurement, and they presently exist in this particular form $|  n_{+} \rangle_{D_{+}} $ $| n_{-}  \rangle_{D_{-}} $ $| n_{H}  \rangle_{D_{H}}$ $ | n_{V} \rangle_{D_{V}}$. Owing to the diverse responses of the detectors, the photons in the signal path $ | m  \rangle_{b_{H}}|n-m\rangle_{b_{V}} $ will undergo a collapse into 16 distinct states.
We postulate that the detection efficiencies of each of the four detectors are $\eta$.
The probabilities with which this state $|  n_{+} \rangle_{D_{+}} $ $| n_{-}  \rangle_{D_{-}} $ $| n_{H}  \rangle_{D_{H}}$ $ | n_{V} \rangle_{D_{V}}$ gives rise to detector responses are as follows respectively:
\begin{eqnarray}
	D_{+}&&=1-(1-Pd)(1-\eta)^{n_{+}},\\ \nonumber
	D_{-}&&=1-(1-Pd)(1-\eta)^{n_{-}},\\\nonumber
	D_{H}&&=1-(1-Pd)(1-\eta)^{n_{H}},\\\nonumber
	D_{V}&&=1-(1-Pd)(1-\eta)^{n_{V}},
\end{eqnarray}
and we define $\bar{D} _{\bullet }=1-D _{\bullet }$ as the probability that detector has no response, where $\bullet$  represents $+$,$-$,$H$, or $V$.

Thus the probability of no detector response can be written as:
\begin{eqnarray}
	\gamma _{0}=\bar{D} _{+ }\bar{D} _{-}\bar{D} _{H}\bar{D} _{V},
\end{eqnarray}
and the state in signal path will collapse into:
 \begin{eqnarray} \label{state0}
	&&| \phi _{0}\rangle =\sum_{n=0}^{\infty } \sum_{m=0}^{n}    \sqrt{P_{0}(m,n-m)}  | m  \rangle_{b_{H}}|n-m\rangle_{b_{V}},
\end{eqnarray}
where
 \begin{eqnarray} \label{pnm0}
&&	P_{0}(m,n-m)= P(m,n-m| P_{0}) =\sum_{n_{H}=n-m-i_{1}-i_{2}} \\\nonumber
&&	\sum_{n_{V}=m-j_{1}-j_{2}}
	\gamma _{0}P_{s}(n_{+},n_{-},n_{H},n_{V},m,n-m),
\end{eqnarray}
which represents the  probability that there exist $n$ photons within the signal path, with $m$ photons being in $|H\rangle$ and $n - m$ photons being in $|V\rangle$.

The probability of one detector response can be written as:
\begin{eqnarray}
	\gamma _{H}=\bar{D} _{+ }\bar{D} _{-}\bar{D} _{H}D _{V},\\
	\gamma _{V}=\bar{D} _{+ }\bar{D} _{-}D_{H}\bar{D}  _{V},\\
	\gamma _{+}=D _{+ }\bar{D} _{-}\bar{D} _{H}\bar{D}  _{V},\\
	\gamma _{-}=\bar{D} _{+ }D _{-}\bar{D} _{H}\bar{D}  _{V}.
\end{eqnarray}
Notice that the responses corresponding to $\gamma _{H}$ and $\gamma _{V}$ are opposite. Due to Eq. (\ref{bell}), when the heralding path is $| H\rangle$ , the signal path will be considered as $| V\rangle$, and vice versa.

In this way the state will collapse into:
\begin{eqnarray} \label{state1}
	&&| \phi _{x}\rangle =\sum_{n=0}^{\infty } \sum_{m=0}^{n}    \sqrt{P_{x}(m,n-m)}  | m  \rangle_{b_{H}}|n-m\rangle_{b_{V}},
\end{eqnarray}
and
 \begin{eqnarray} \label{pnm1}
	P_{x}(m,n-m)=&&
	\sum_{n_{H}=n-m-i_{1}-i_{2}} \sum_{n_{V}=m-j_{1}-j_{2}}  \\\nonumber
	&&\gamma _{x}P_{s}(n_{+},n_{-},n_{H},n_{V},m,n-m),
\end{eqnarray}
where x correspond to H, V,+,-.

In total, there exist 16 kinds of detector responses. Based on these responses, we can similarly formulate the corresponding response probabilities as:
 \begin{eqnarray}
	\gamma _{x}=\bar{D} _{\bar{x} }D _{x}.
\end{eqnarray}
For example, when the $D_{H}$ and $D_{+}$ responses occur, corresponding to $x=H+$, $\gamma _{x}$ can be written as $\bar{D} _{V}\bar{D} _{-}D _{H}D _{+ }$.
When x is $HV+$, $\gamma _{x}$ will be $\bar{D} _{-}D _{H}D _{V}D _{+ }$.

The states in signal path and the corresponding probability will be represented similarly as Eq. (\ref{state1}) and Eq. (\ref{pnm1}).

In a word, when no response occurs, $x=0$.
When there is a detector response, $x\in \{H,V,+,-\}$, which are used to generate key.
When there are two detectors response, $x\in \{HV,+-,H+,H-,V+,V-\}$.
When there are three detectors response, $x\in \{HV+,HV-,V+-,H+-\}$.
When  all four detectors respond, $x=4$.

\subsection{The channel model}
According to the derivation in the previous section, the form of photons entering the channel is $| m  \rangle_{b_{H}}|n-m\rangle_{b_{V}}$. Depending on the different responses in heralding path, they possess different probability distributions $P_{x}(m,n-m)$.
Suppose the channel attenuation is $\eta_{c}=10^{-\alpha L/10}$, where $\alpha$ is fiber attenuation coefficient and L is the transmission distance. Bob conducts a Z-basis measurement (the case of X-basis is discussed in the appendix B). The detection efficiency of the detector at Bob's end is $\eta_{D}$.
Then the yield of the state $| m  \rangle_{b_{H}}|n-m\rangle_{b_{V}}$ can be calculated as:
\begin{eqnarray}
	Y_{m,n-m}=&&D_{H}^{B} (1-D_{V}^{B})+D_{V}^{B}(1-D_{H}^{B}),
%	=&&[1-(1-Pd)(1-\eta_{c}\eta_{D})^{m}](1-Pd)(1-\eta_{c}\eta_{D})^{n-m}+\\\nonumber
%	&&[1-(1-Pd)(1-\eta_{c}\eta_{D})^{n-m}](1-Pd)(1-\eta_{c}\eta_{D})^{m}\\\nonumber
\end{eqnarray}
where
\begin{eqnarray}
D_{H}^{B}&&=1-(1-Pd)(1-\eta_{c}\eta_{D})^{m}, \\
D_{V}^{B}&&=1-(1-Pd)(1-\eta_{c}\eta_{D})^{n-m}.
\end{eqnarray}
$D_{H}^{B}$ ($D_{V}^{B}$) represents the detector $D_{H}$ ($D_{V}$) of Bob has a response, and Pd is the dark count rate.

Taking the $|   H \rangle$ state as an example, the error rate can be calculated as:
\begin{eqnarray}
	e_{m,n-m}Y_{m,n-m}=&&e_{d}D_{H}^{B} (1-D_{V}^{B})+(1-e_{d})D_{V}^{B}(1-D_{H}^{B}), \nonumber \\
	e_{m,n-m}=&&\frac{e_{m,n-m}Y_{m,n-m}}{Y_{m,n-m}} ,
\end{eqnarray}
where $e_{d}$ is basis misalignment error rate.

\subsection{Key generation rate}
In our model, given that the Z-basis and X-basis are generated with equal probabilities (it should be noted that modifying the transmission and reflection ratio of the BS in Fig 1 can adjust the probability of basis selection), we postulate that both the Z-basis and X-basis are employed for key generation. And the key generation rate can be written as:
\begin{eqnarray}
	R=q[P_{1}Y_{1}(1-h(e_{1})-Qfh(E)]
\end{eqnarray}
where $q$ is the basis reconciliation factor and in our protocol, it is $ \frac{1}{2} $. $P_{1}Y_{1}$ correspond to single-photon gain and $e_{1}$ is the single-photon error rate.
Q is the overall gain and E is the bit error rate, which can  be modeled as:
\begin{eqnarray}
	Q_{x}&&=\sum_{n=0}^{\infty } \sum_{m=0}^{n} P_{x}(m,n-m)Y_{m,n-m}\\
	Q_{x}E_{x}&&=\sum_{n=0}^{\infty } \sum_{m=0}^{n} P_{x}(m,n-m)Y_{m,n-m}e_{m,n-m}\\
	E_{x}&&=\frac{Q_{x}E_{x}}{Q_{x}}
\end{eqnarray}

\subsection{Decoy state analysis}
In the field of QKD, the decoy state method is utilized to estimate the single-photon yield and error rate by means of the total gain and bit error rate of light sources with diverse distributions. In our herald fully passive setup, we need to estimate $P_{1}Y_{1}$ and $e_{1}$, which can be obtained by solving the following linear programming problems:
\begin{eqnarray}
	min  && \quad P_{H}(1,0)Y_{1,0}+ P_{H}(0,1)Y_{0,1}   \\ \nonumber
	s.t.&&\quad Q_{x}   \ge \sum_{n=0}^{n_{cut}}\sum_{m=0}^{n}   P_{x}(m,n-m)   Y_{m,n-m}   ,\\  \nonumber
	&&\quad Q_{x}   \le \sum_{n=0}^{n_{cut}} \sum_{m=0}^{n}   P_{x}(m,n-m)   Y_{m,n-m}   \\  \nonumber
&&\quad	\quad\quad\quad +1-\sum_{n=0}^{n_{cut}} \sum_{m=0}^{n} P_{x}(m,n-m)   ,\\  \nonumber
	&&\quad 0\le  Y_{m,n-m}     \le 1 ,
\end{eqnarray}
\begin{eqnarray}
	max  && \quad P_{H}(1,0)Y_{1,0}e_{1,0} + P_{H}(0,1)Y_{0,1}e_{0,1}     \\ \nonumber
	s.t.&&\quad Q_{x} E_{x}  \ge \sum_{n=0}^{n_{cut}}\sum_{m=0}^{n}   P_{x}(m,n-m)   Y_{m,n-m} e_{m,n-m}  ,\\  \nonumber
	&&\quad Q_{x}E_{x}    \le \sum_{n=0}^{n_{cut}} \sum_{m=0}^{n}   P_{x}(m,n-m)   Y_{m,n-m} e_{m,n-m}  \\  \nonumber
	&&\quad	\quad\quad\quad +1-\sum_{n=0}^{n_{cut}} \sum_{m=0}^{n} P_{x}(m,n-m)   ,\\  \nonumber
	&&\quad 0\le  Y_{m,n-m} e_{m,n-m}    \le 1 ,
\end{eqnarray}
where x represents the 16 kinds of detector responses described in section \ref{sa}, and the two linear programming problems are designed for resolving the parameter of $| H \rangle$ state.
Then, $P_{1}Y_{1}$ and $e_{1}$ can be calculated as:
\begin{eqnarray}
	P_{1}Y_{1} &&=P_{H}(1,0)Y_{1,0}+ P_{H}(0,1)Y_{0,1},\\
	e_{1}&&=\frac{P_{H}(1,0)Y_{1,0}e_{1,0} + P_{H}(0,1)Y_{0,1}e_{0,1}}{P_{1}Y_{1}} .
\end{eqnarray}
Owing to symmetry, the circumstances of other states are analogous.

\section{Numerical simulation}

In this section, we perform numerical simulation of our herald fully passive QKD protocol.
In the numerical simulation, the parameter settings are as follows: The detector efficiencies, $\eta$ and $\eta_{D}$, are both set at 0.65. The dark count rate, Pd, is configured to be $10^{-6}$ per pulse. The fiber attenuation coefficient is set to 0.2 dB/km. The detector basis misalignment error rate, $e_{d}$, is set to 0.015. And the basis  reconciliation factor, q, is $ \frac{1}{2} $.
\begin{table}[h]
	\centering
	\vspace{-0.1cm}%调整图片与上下文的垂直距离
	\setlength{\abovecaptionskip}{0.3cm}%调整图片标题与图距离
	\setlength{\belowcaptionskip}{0.1cm}%调整图片标题与下文距离
	\setlength\tabcolsep{5pt} %调整表格列宽度
	\renewcommand\arraystretch{1.5}  %调整表格行间距
	\caption{Parameters used in the numerical simulation.}
	\begin{tabular}{cccccc}
		\hline
		\hline
		$\eta$ & $\eta_{D}$ & Pd &$e_{d}$   & $\alpha$   & $q$\\
		0.65    & 0.65    & $10^{-6}$  &0.015    & 0.2    &  $ \frac{1}{2} $ \\  \hline \hline
	\end{tabular}
\end{table}

\begin{figure}[!htbp]%[tpb]
	\begin{center}
		\includegraphics[width=9cm,angle=0]{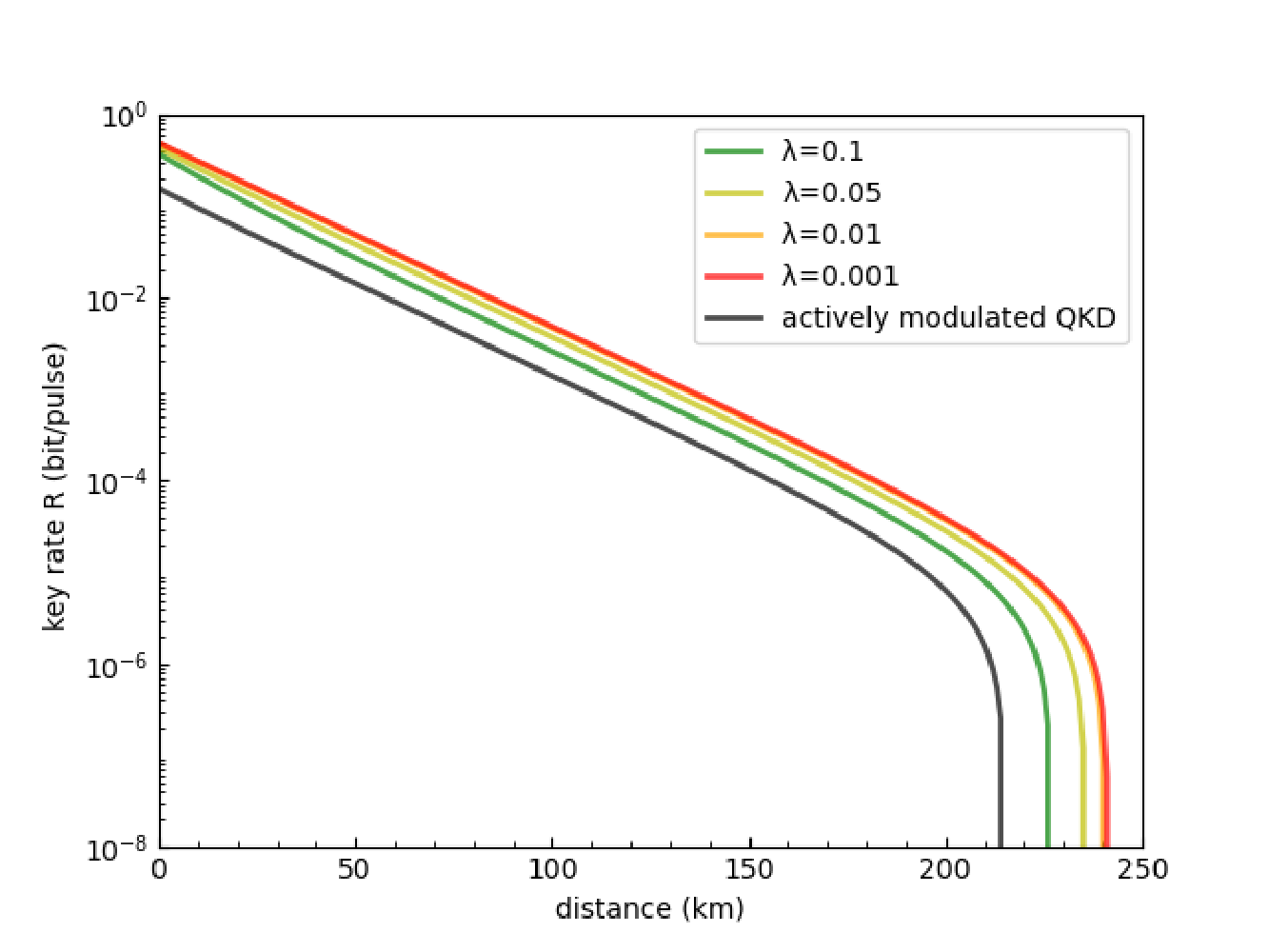}
		%\captionsetup{justification=raggedright}
		\caption{Comparison of the key rate of the herald fully passive QKD with different $\lambda$. The red, orange, yellow, and green curves  correspond to herald fully passive QKD with $\lambda=0.001,0.01,0.05,0.1$, respectively. The black curve represents the optimal actively modulated QKD. Notice that the red curve and the orange curve almost overlap.
		}\label{bpb}
	\end{center}
\end{figure}

In Fig. \ref{bpb}, we present the key rate curves for different values of $\lambda$, where $2\lambda$ represents the average number of photon pairs.
The red, orange, yellow, and green curves  correspond to herald fully passive QKD with $\lambda=0.001,0.01,0.05,0.1$, respectively.
As is observable from Fig. \ref{bpb}, these curves exhibit a high degree of proximity. The herald operation is capable of eliminating a substantial portion of the vacuum state component within the light source. This, in turn, augments the proportion of single-photon component, leading to a remarkable enhancement in the key rate and an elongation of the transmission distance. In comparison with the actively modulated QKD without herald (the black curve), the key rate of the heralded fully passive QKD has been elevated to over three times of that of the actively modulated QKD, and the transmission distance has been extended to 241 kilometers.

Moreover, it can be noted that as $\lambda$ diminishes from 0.1 to 0.001, the overall performance of the protocol, in terms of both the key rate and the transmission distance, continues to ascend. The rationale behind this lies in the fact that a smaller $\lambda$ implies a lesser presence of multi-photon component within the light source. This serves to effectively curtail the amount of key information that could potentially be intercepted by an eavesdropper. Concurrently, the herald operation plays a crucial role in maintaining a high  proportion of single photons. Hence, from the perspective of pulse performance, a smaller $\lambda$  is generally more favorable. Nevertheless, it should be borne in mind that a decrease in $\lambda$  also leads to a reduction in the success rate of herald. Although we have utilized all the herald response events, only the single-photon responses can effectively generate keys. This, unfortunately, can result in a significant consumption of resources and subsequently lead to a decline in the key transmission rate.  Therefore, selecting an appropriate $\lambda$ is important for the practical application of the protocol.

\begin{figure}[!htbp]%[tpb]
	\begin{center}
		\includegraphics[width=9cm,angle=0]{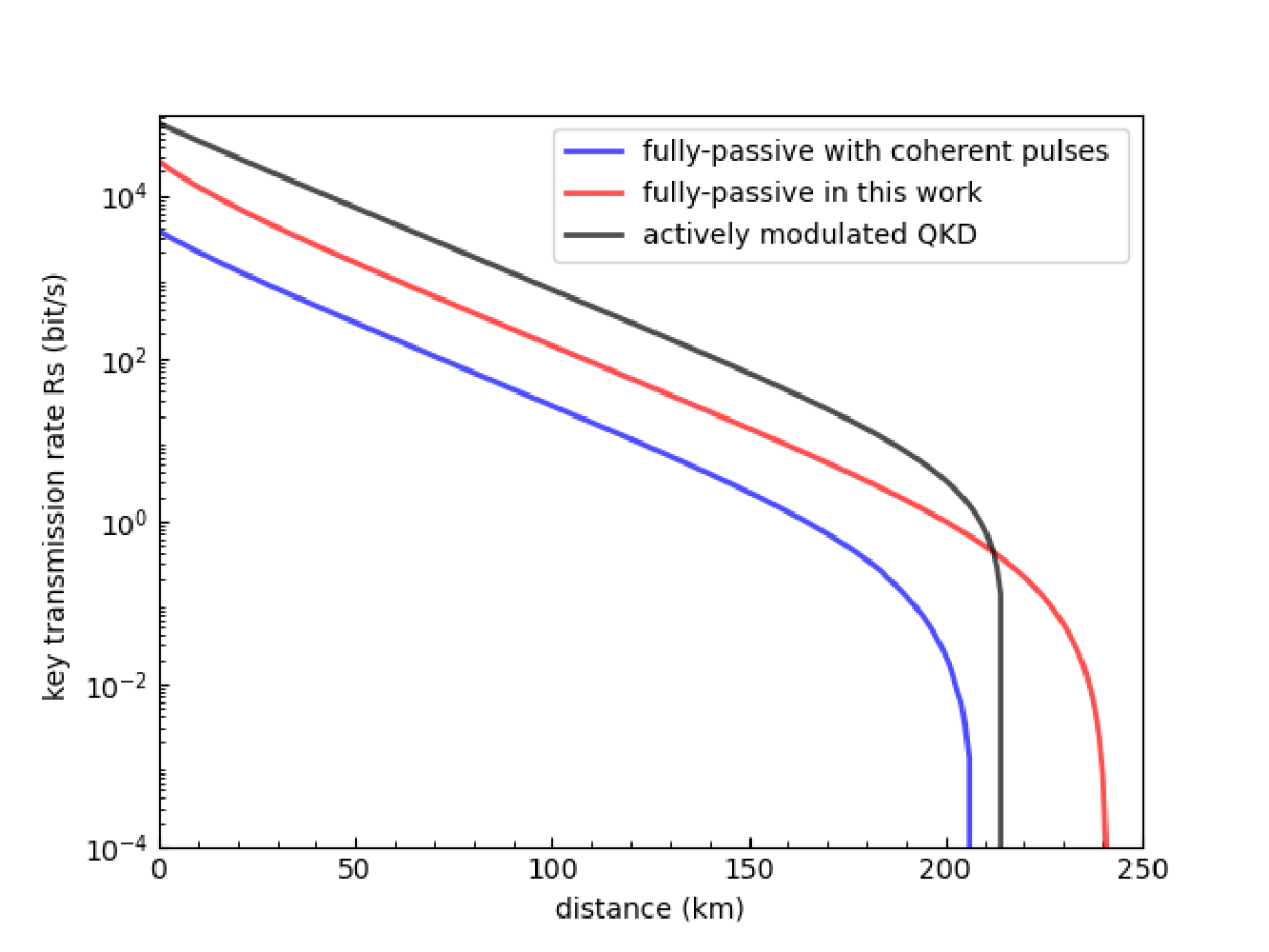}
		%\captionsetup{justification=raggedright}
		\caption{Comparison of the optimal  key transmission rate of the herald fully passive QKD with PDC, actively modulated QKD, and the fully passive QKD with phase-randomized coherent pulses. The black, red, blue curves represent the performance of  actively modulated QKD, herald fully passive QKD with PDC, and the fully passive QKD with phase-randomized coherent pulses \cite{FP3}, respectively.
		}\label{bps}
	\end{center}
\end{figure}

In Fig. \ref{bps}, a practical laser source is taken into consideration, whose frequency is configured at $10^{6}$ Hz. Under such circumstances, we take into account the herald success rate and optimize the key transmission rate of the actual protocol. The red curve in Fig. \ref{bps} represents the optimal performance curve of the fully passive QKD proposed by us. The black curve corresponds to the optimal performance curve of the actively modulated QKD. And the blue curve depicts the optimal performance curve of the fully passive QKD realized by means of phase-randomized coherent pulses \cite{FP3}.

\begin{figure}[!htbp]%[tpb]
	\begin{center}
		\includegraphics[width=9cm,angle=0]{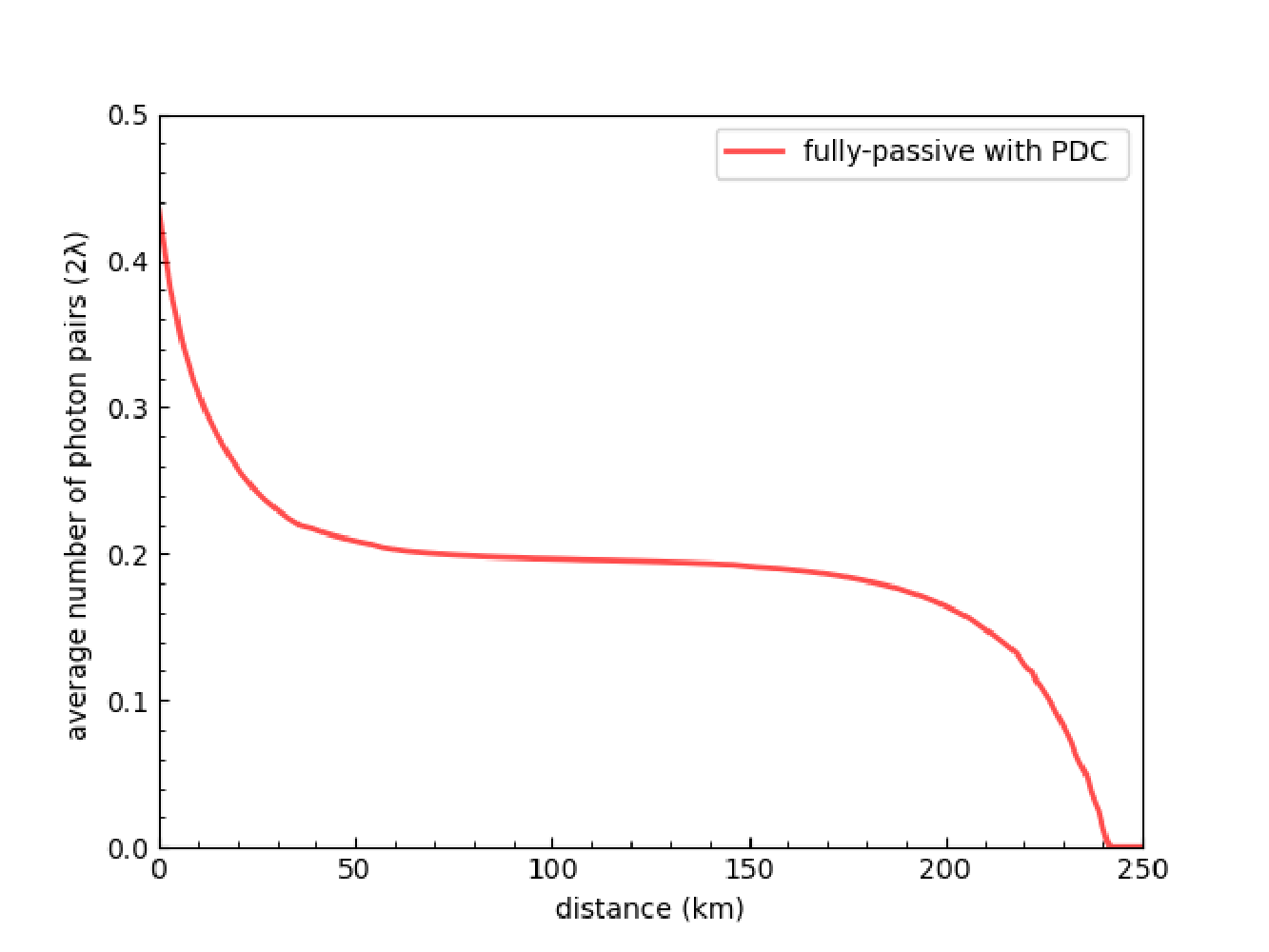}
		%\captionsetup{justification=raggedright}
		\caption{The optimal average number of photon pairs (2$\lambda$) corresponding to our protocol in Fig. \ref{bps}.
		}\label{mu}
	\end{center}
\end{figure}

Among them, both of the two fully passive protocols require the consumption of certain resources to achieve passive encoding and passive decoy states. In the case of the scheme using coherent light for realizing full passive, its passive decoy states will augment the multi-photon components of the light source and thereby reduce the key rate of the protocol. In addition, owing to the randomness inherent in passive encoding, a substantial quantity of states will be discarded during post-selection, and additionally, some imperfect states will be generated. These two aspects exert an influence on its key transmission rate.
In contrast, our protocol accomplishes passive encoding and passive decoy states simultaneously while performing herald operation. The principal factor affecting the key transmission rate herein is mainly the probability of herald single photons. Moreover, as a result of the herald operation, the quality of our light source is considerably superior to that of the actively modulated QKD, which enables a significant enhancement in the transmission distance.
The optimal average photon pair number $2\lambda$ of our protocol  in Fig. \ref{bps} is illustrated in Fig. \ref{mu}.

\begin{figure}[!htbp]%[tpb]
	\begin{center}
		\includegraphics[width=9cm,angle=0]{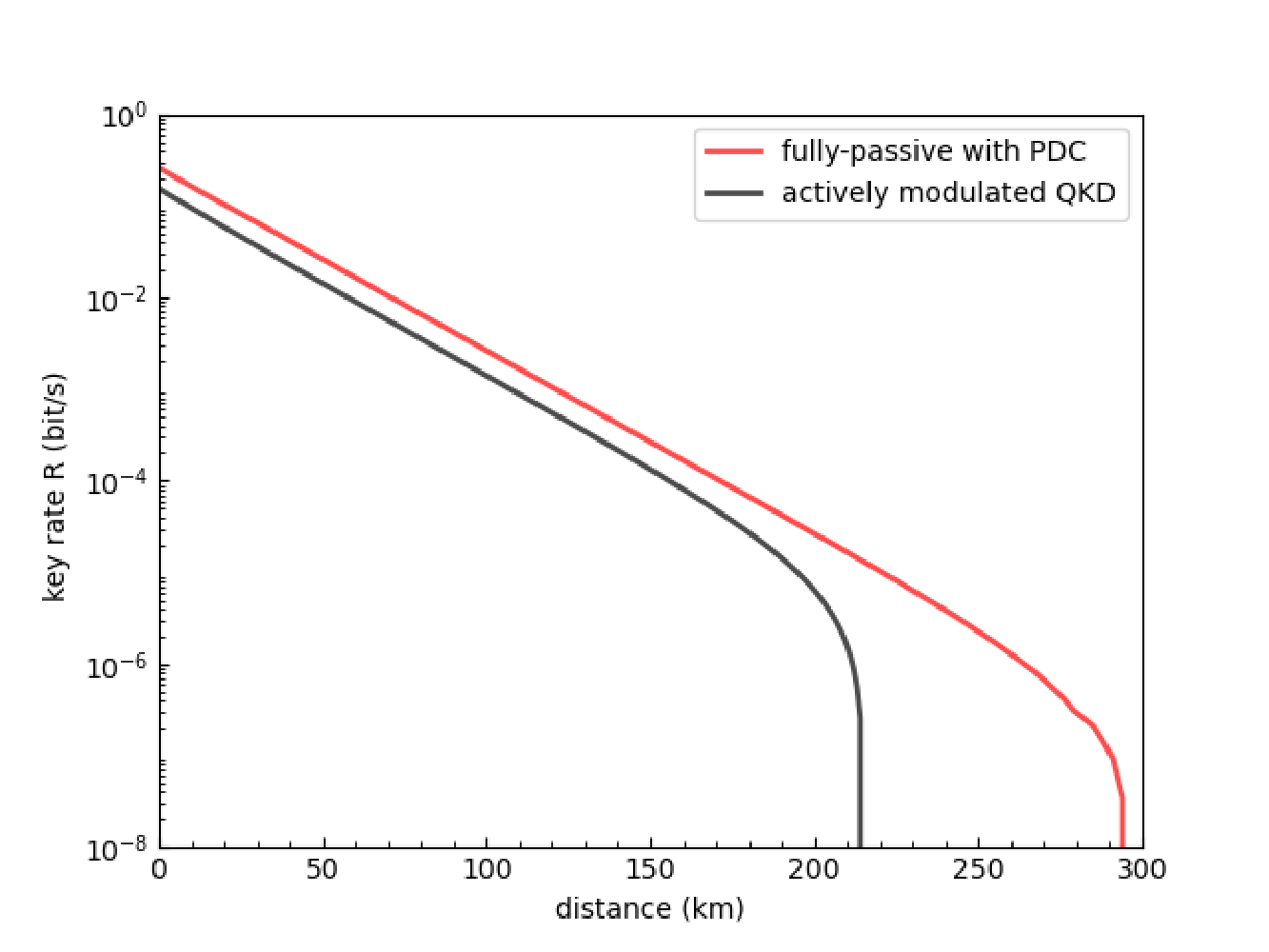}
		%\captionsetup{justification=raggedright}
		\caption{ The performance of the pulse when the maximum transmission rate is achieved, corresponding to the average number of photon pairs in Fig. \ref{mu}.
		}\label{com}
	\end{center}
\end{figure}

In Fig. \ref{com},  the key rate of our protocol is presented under the circumstance of attaining the maximum transmission rate. In comparison with the actively modulated QKD, its key rate still exhibits a 70\% enhancement.

\section{Conclusion}
In this paper, we introduce a herald fully passive QKD protocol based on PDC. By utilizing the heralding operation, we simultaneously generate decoy states and perform passive encoding, eliminating the need for active modulation procedures. This approach not only simplifies the system architecture but also enhances resistance to side-channel attacks by third-party eavesdroppers targeting the source modulator. Owing to the heralding operation, the key rate of our protocol exceeds that of actively modulated QKD by more than threefold, while the communication distance is expanded to 241 km. Furthermore, we account for actual resource consumption and optimize the key transmission rate of our protocol. In comparison to fully passive QKD protocols that rely on coherent pulses, our approach demonstrates significant advantages in both key rate and key transmission rate.

\section{Appendix A: the passive decoy state method with PDC}
In this section, we will briefly introduce the passive decoy state method based on Parametric Down-Conversion (PDC), enabling readers to understand our work more easily.

First, let us introduce the concept of the decoy state method. In an ideal quantum communication protocol, a single-photon source is desired. However, in practice, the photons emitted from the light source often deviate from this ideal state, generating vacuum states, single-photon states, or multiple-photon states. In such cases, an eavesdropper can perform PNS attack on the multi-photon components in the channel to capture all keys associated with those photons, while simultaneously executing a coherent attack on the single-photon components, extracting keys from the errors in the single photons. This naturally raises the question: how many secure keys can ultimately be obtained?
In 2004, GLLP theory was proposed to establish a system model for practical weak coherent light sources, enabling the estimation of the secure key rate \cite{GLLP}. In 2005, Lo, and Ma, et.al. integrated the GLLP theory with the decoy state method \cite{decoy,decoy1}, allowing for a more precise estimation of the system's error rate and yield, thereby enhancing the system performance. The essence of the decoy state method lies in addressing a linear programming problem using multiple light sources with different distributions; the greater the number of light sources employed, the more accurate the resulting solution.

\begin{figure}[!htbp]%[tpb]
	\begin{center}
		\includegraphics[width=9cm,angle=0]{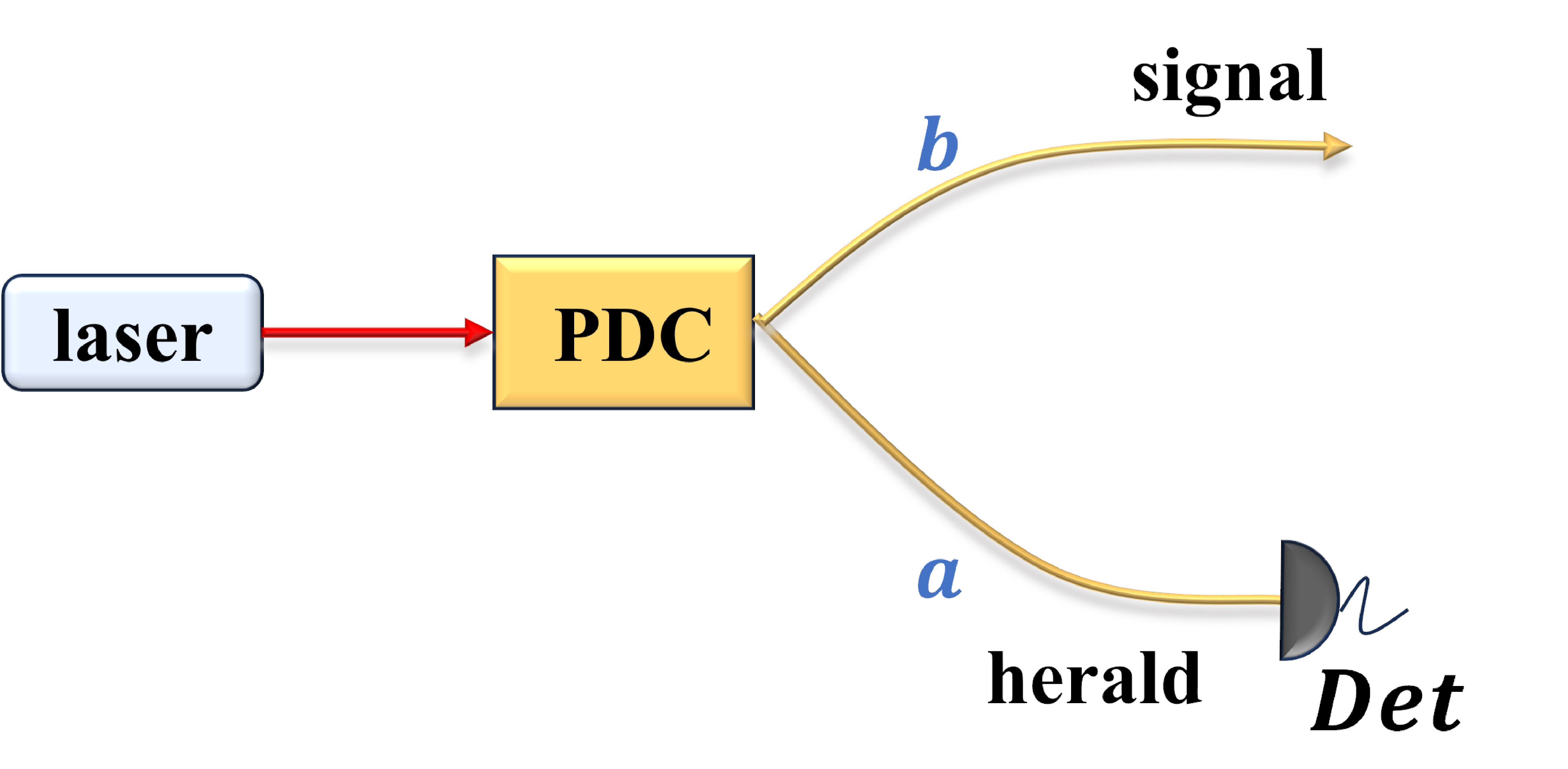}
		%\captionsetup{justification=raggedright}
		\caption{ The structure of passive decoy state source \cite{pdecoypdc1}.
		}\label{pPDC}
	\end{center}
\end{figure}

The commonly used method for generating different sources is to modulate the intensity of the laser source.
Another approach is to classify optical pulses through post-measurement selection, which is known as the passive source scheme. There are two types of passive source schemes \cite{pdecoy,pdecoypdc1}, and here we mainly introduce the scheme based on Parametric Down-Conversion (PDC). The structure of passive decoy state source is shown in Fig. \ref{pPDC} \cite{pdecoypdc1}.

When a laser is incident on a PDC crystal, photon pairs are generated. The generated states can be written as \cite{pdecoypdc1}:
\begin{eqnarray}
	| \psi   \rangle &&=\sum_{n=0}^{\infty } \frac{\lambda ^{n}}{\sqrt{n!} } e^{-\lambda }
	| n   \rangle_{s}| n   \rangle_{h}, \\
	P(n)&&=\frac{ | \lambda   |^{2n} }{n!}e^{-2\lambda }
\end{eqnarray}

The n photons on the heralding path may or may not cause the detector to respond.  The probabilities of them can be written as:
\begin{eqnarray}
	P_{D}&&=1-(1-pd)*(1-\eta )^{n},\\
	P_{ND}&&=(1-pd)*(1-\eta )^{n},
\end{eqnarray}
where pd is the dark count rate, and $\eta$ is the detector efficiency.

According to whether the detector responds or not, we can divide the photons on the heralding path into two categories as:
\begin{eqnarray}
	P(n |  P_{D}) &&= P(n)*P_{D},\\
	P(n |  P_{ND}) &&= P(n)*P_{ND},
\end{eqnarray}
and they correspond to the state:
\begin{eqnarray}
	| \psi   \rangle_{D}&&=\sum_{n=0}^{\infty } \sqrt{P(n |  P_{D})}| n   \rangle_{s}\\
	| \psi   \rangle_{ND}&&=\sum_{n=0}^{\infty } \sqrt{P(n |  P_{ND})}| n   \rangle_{s}
\end{eqnarray}

Through this method,   two different source with different photon number distributions can be generated passively without modulating the intensity of the source.

\section{Appendix B: the case of X-basis}

When Bob conduct a X-basis measurement, the state $| \phi _{x}\rangle$  can be projected onto the X basis form as:
\begin{eqnarray}
	| \phi _{x}\rangle &&=\sum_{n=0}^{\infty } \sum_{m=0}^{n}    \sqrt{P_{x}(m,n-m)}  | m  \rangle_{b_{H}}|n-m\rangle_{b_{V}}\\ \nonumber
	&&=\sum_{n=0}^{\infty } \sum_{m=0}^{n}    \sqrt{P_{x}(m,n-m)}\frac{1}{\sqrt{m!(n-m)!} } \\\nonumber
	&& \quad (\frac{1}{\sqrt{2} } )^{n}(b_{+}^{\dagger }+b_{-}^{\dagger })^{m}
	(b_{+}^{\dagger }-b_{-}^{\dagger })^{n-m}\\\nonumber
	&&=\sum_{n=0}^{\infty } \sum_{m=0}^{n}    \sqrt{P_{x}(m,n-m)}\frac{1}{\sqrt{m!(n-m)!} }(\frac{1}{\sqrt{2} } )^{n} \\\nonumber
	&& \quad \sum_{x=0}^{m}\sum_{y=0}^{n-m}
	C_{m}^{x} C_{n-m}^{y} (-1)^{n-m-y}(b_{+}^{\dagger })^{x+y}(b_{-}^{\dagger })^{n-(x+y)}\\\nonumber
	&&=\sum_{n=0}^{\infty } \sum_{m=0}^{n}    \sqrt{P_{x}(m,n-m)}\frac{1}{\sqrt{m!(n-m)!} }(\frac{1}{\sqrt{2} } )^{n} \\\nonumber
	&&\quad \sum_{t=0}^{n} \sum_{x+y=t}C_{m}^{x} C_{n-m}^{y} (-1)^{n-m-y}(b_{+}^{\dagger })^{t}(b_{-}^{\dagger })^{n-t}\\\nonumber
	&&=\sum_{n=0}^{\infty } \sum_{m=0}^{n}    \sqrt{P_{x}(m,n-m)}\frac{1}{\sqrt{m!(n-m)!} }(\frac{1}{\sqrt{2} } )^{n} \\\nonumber
	&&\quad \sum_{t=0}^{n} \sum_{x+y=t}C_{m}^{x} C_{n-m}^{y} (-1)^{n-m-y}\sqrt{t!(n-t)!} | t  \rangle_{b_{+}}|n-t\rangle_{b_{-}}\\\nonumber
	&&=\sum_{n=0}^{\infty }\sum_{t=0}^{n}\sqrt{P_{x}^{X}(t,n-t)}| t  \rangle_{b_{+}}|n-t\rangle_{b_{-}}
\end{eqnarray}
where
\begin{eqnarray}
	&&P_{x}^{X}(t,n-t)=|\sum_{m=0}^{n} \sum_{x+y=t}\sqrt{P_{x}(m,n-m)}\\ \nonumber
	&&\frac{\sqrt{t!(n-t)!}}{\sqrt{m!(n-m)!} }(\frac{1}{\sqrt{2} } )^{n}C_{m}^{x} C_{n-m}^{y} (-1)^{n-m-y}|^{2}
\end{eqnarray}
In this way, the form of photons entering the channel is $| t  \rangle_{b_{+}}|n-t\rangle_{b_{-}}$. Depending on the different responses in heralding path, they possess different probability distributions $P_{x}^{X}(t,n-t)$. The subsequent process is similar to that in the main text.

\section*{Acknowledgement}
 This work is supported by the National Natural Science Foundation of China under Grants  No. 12175106 and  No. 92365110.

\end{document}